Irradiation Facilities and Irradiation Methods for High Power Target

# FERMILAB-CONF-22-124-AD

F. Pellemoine [1,*], K. Ammigan [1], S. Bidhar [1], B. Zwaska [1]
C. Barbier [2], D. McClintock [2], S. Taller [2], D. Winder [2]
Y. Sun [3]
C. S. Cutler [4], D. Kim [4]
Y. Chiu [5], M. Freer [5], C. Wheldon [5]
A. Gottberg [6], F. Boix Pamies [6]
M. Calviani [7], N. Charitonidis [7], R. Garcia Alia [7], F. Ravotti [7], S. Danzeca [7], A.P. Bernardes [7]
N. Moncoffre [8]
S. Meigo [9], T. Ishida [9]
Y. Dai [10]
A. Couet [11], K. Kriewaldt [11]
M. Moorhead [12]
G.S. Was [13], O. Toader [13], F. Naab [13], P. Wang [13], D. Woodley [13]
E. Getto [14]
S. Raiman [15]
C. Grygiel [16], I. Monnet [16]
A. Alessi [17]
D.J. Senor, A.M. Casella [18]

1) Fermi National Accelerator Laboratory, Batavia IL 60510-5011 USA
2) Oak Ridge National Laboratory, USA
3) Argonne National Laboratory Argonne IL 60439, USA
4) Brookhaven National Laboratory, USA
5) University of Birmingham, UK
6) TRIUMF and University of Victoria, Canada
7) CERN, Switzerland
8) Univ Lyon, Univ Claude Bernard Lyon 1, CNRS/IN2P3, IP2I Lyon, F-69622, Villeurbanne, France
9) J-PARC, KEK
10) Paul Scherrer Institut, Switzerland
11) University of Wisconsin – Madison, Madison, WI 53715, USA
12) Idaho National Laboratory, USA
13) University of Michigan, USA
14) United States Naval Academy, USA
15) Texas A&M University, USA
16) CIMAP - Grand Accelerateur National d'Ions Lourds, 14076 CAEN Cedex 5, France
17) LSI, CEA/DRF/IRAMIS, CNRS, Ecole polytechnique, Institut Polytechnique de Paris, 91120 Palaiseau, France
18) Pacific Northwest National Laboratory, Richland, WA, USA

*Corresponding author: fpellemo@fnal.gov*

Abstract

High Power Target systems are key elements in future neutrino and other rare particle production in accelerators. These systems transform an intense source of protons into secondary particles of interest to enable new scientific discoveries. As beam intensity and energies increase, target systems face significant challenges. Radiation damages and thermal shocks in target materials were identified as the leading cross-cutting challenges of high-power target facilities. Target material R&D to address these challenges are essential to enable and ensure reliable operation of future-generation accelerators. Irradiation facilities and alternative methods are critical to provide a full support of material R&D and better address these critical challenges.

# 1 INTRODUCTION

In the recent past, several accelerator facilities have had to limit their beam powers, not as a result of limitations of the accelerators themselves, but due to the target and/or window and their survivability. The Neutrinos at the Main Injector (NuMI) beamline at Fermilab [1], the Materials and Life Science Experimental Facility (MLF) at J-PARC [2] and the Spallation Neutron Source (SNS) at Oak Ridge National Laboratory [3] have operated at reduced power for extended periods due to target failure concerns. Recent beam power upgrades at various accelerator facilities are now pushing conventional materials to their limits, and future facilities will present even greater challenges in meeting the physics goals of these experiments.

Beam-intercepting materials that are resilient and capable of withstanding an order of magnitude increase in particle beam intensities are essential for enabling next-generation multi-megawatt target accelerator facilities. It is now critical to explore novel targets and advanced materials beyond the current state-of-the-art to improve resistance to beam-induced radiation damage and thermal shock - the leading cross-cutting challenges facing high-power target facilities [4] for current and next-generation accelerators. Conventional materials for key accelerator components, such as beam windows and secondary particle production targets, are already limiting the scope of experiments. Novel target materials will enable multi-MW beam operation to fully satisfy design requirements and maximize the physics benefits of future High Energy Physics (HEP) experiments.

The development of novel materials and concepts [5] for future -accelerators is mainly supported by simulations at different levels: prediction of beam-matter interactions, thermo-mechanical response induced by beam heating, evaluation of radiation damage with time that enhances degradation of material physical properties. As the next generation of high-power targets will use more complex geometries, novel materials and concepts, the current numerical approaches will not be sufficient and advanced modeling needs to be developed [6]. In order to obtain more confidence in the numerical results to develop more reliable target systems, better validation must be performed in parallel to the simulation development by using prototypes tested under relevant beam conditions.  Although some data are available in the literature, more experimental data with controlled beam parameters are needed for targets operating under extreme conditions where the boundary conditions are not fully controlled. Fully instrumented targets with simple geometry are usually difficult to insert online during accelerator operations as they are not dedicated for materials research. Existing test facilities that can reproduce similar beam conditions that accelerator operation parameters are very limited or don't exist.

Exposure of material to particles leads to a displacement of a significant number of atoms from their preferential lattice (quantified as displacements per atom, DPA, a reference used to express radiation damage level in material), transmutation and gas production. Irradiation-induced defects such as dislocation loops, point defect clusters, fine-scale precipitates and voids that accumulate at the microstructural level ultimately disrupt the lattice structure of the material and affect the mechanical, electrical and other physical properties of irradiated materials. Typical bulk material effects include embrittlement, hardening, swelling, reduction of thermal conductivity, and an increase in diffusion-dependent phenomena such as segregation of impurities, and phase transformation [7], all of which are detrimental to the thermo-mechanical health and physics performance of the material.

Thermal shock phenomena arise in beam-intercepting materials as a result of localized energy deposition caused by very short pulsed-beams (1-10 µs). The typical peak temperature increase in a 1-MW neutrino target at Fermilab is approximately 250 K in 10 µs ($2.5 \times 10^7$ K/s). The rapidly heated core volume of the material in the beam spot region expands but is constrained by the surrounding cooler target material. This condition creates a sudden localized region of compressive stress that propagates as stress waves through the material at sonic velocities. If the initial dynamic stress exceeds the yield strength of the material, it will permanently deform and eventually fail. In addition, the cyclic loading environment from the pulsed beam progressively damages the materials' microstructure such that it ultimately fails at stress levels that are lower than its failure strength (fatigue failure).

The pulsed beam structure and small beam spot size (~1 mm) in accelerators also induce more severe thermal shock effects. As a result, materials radiation damage data that cover the accelerator irradiation parameter space are lacking in the literature. It is therefore imperative to understand the behavior of materials under irradiation conditions analogous to future accelerator target facilities.

With present plans to upgrade accelerator facilities at Fermi National Accelerator Laboratory (FNAL) to higher beam powers (1.2+ MW) in the next decade, timely R&D of robust high-power targets is essential to fully secure the physics benefits of ambitious accelerator power upgrades.

A lack of irradiation facilities dedicated to material science to better understand radiation damage for material relevant for accelerator targets slows down this effort.

Since 2013, the Radiation Damage In Accelerator Target Environments (RaDIATE) collaboration [8] has been addressing these material challenges by engaging experts from the nuclear materials and accelerator target communities to perform mutually beneficial research [9] and collect experimental data from known beam parameters. The collaboration consists of over 70 participants from 14 national and international institutions working together on various accelerator materials R&D projects. This highlights the importance and concerns in meeting the demands of future accelerator target facilities.

This white paper presents the needs in the next 10 years to support modeling, material development and design improvement for current and future high-power targets at a Multi-MegaWatt beam level. This discussion is the results of the Letter of Interest collected and the workshops organized for Snowmass21.

## 2  Alternative beams

Here we propose a list of different beams that could simulate radiation damage, thermal shock in material or simulate similar heat deposition in prototypes to validate novel design.

## 2.1 Alternative for radiation damage

Directly studying material properties of targets exposed to high energy proton accelerators is an expensive and time-consuming method due to the high level of activation and the low radiation damage rate. The high activation of samples requires characterization in hot cells equipped with advanced test machines and the high user costs/restrictive beam time availability associated with these facilities.

Irradiation with neutrons, heavy ions and electrons cannot well simulate that of high energy protons, because of far less or no transmutant production. In materials irradiated in spallation targets, the radiation effect in materials is primarily displacement damage in low dose cases. Helium effects become pronounced when its concentration is above 500 appm. Solid transmutants may also form high-density nano-clusters and segregation at grain boundaries, although the effects of the clusters and grain boundary segregation are still not understood. Heavy ion irradiation may be used to study displacement damage effect of neutrons, and protons particularly at low doses for fundamental understanding.

The damage accumulation rate (DPA/s), DPA dose concentration/gradients and helium gas production (He appm/DPA) are orders of magnitude higher in accelerator beam-intercepting materials than in neutron-irradiated nuclear reactor materials [10]. Helium production in accelerator systems may have a significant impact in material physical property changes and must be carefully studied.

Existing facilities with adjustments can provide environments close to displacement rate and gas production created by high energy high intensity proton beam. Future developments are obviously needed to validate that radiation damage creation and associated changes in physical properties are similar despite the radiation sources.

**Low Energy Proton Beams** – Lowering the energy of proton beams increases significantly the damage rate, reduces the activation in materials and reduces costs and the time of the irradiation experiment. Proton beams with energies of the order of 10's of MeV could irradiate bulk material (few mm) and therefore allow for conventional thermal and mechanical tests. Handling activated materials still needs long hand-tools or Post-Irradiation Examination (PIE) features such as hot cells.

**Neutron Irradiations** - In fission reactor neutron irradiation tests, helium is rarely produced, but in fusion reactor neutron irradiation fields, the evaluation of the effects of helium produced by nuclear transmutation is an important issue. Radiation damage with neutrons to compare with high energy protons might be investigated as a large amount of experimental data are published for several materials of interest for target systems. Handling highly activated materials will still need Post-Irradiation Examination features such as hot cells with limited thermo-mechanical tests.

**Low Energy Heavy Ion beams** - Indeed, ion irradiation has long been used as a surrogate for neutron irradiation experiments which are comparatively slower, more expensive, and often can leave materials activated and difficult to handle post-irradiation.

Using low energy (LE) heavy ion irradiations could possibly address these issues and limitations, providing an effective and fast way to produce radiation damage in materials without activating the specimens. It also enables effective exploration of candidate materials for future high intensity target facilities. Low energy heavy ion beams provide a much higher damage rate than high energy proton beams: up to few DPA in a few days compared to less than 1 DPA after ~10 weeks of high-energy proton irradiation at the Brookhaven Linac Isotope Producer (BLIP). One drawback is the shallow damage depth (between 1 to a few micrometers depending on ion energy and material) from LE ions. However, micromechanical testing techniques can effectively probe the shallow damage region to measure relative changes in thermal and mechanical properties before and after irradiation. The nanoindentation technique makes it easy to investigate the effect of irradiation on the hardness (proportional to the yield strength) of materials. It should be noted, however, that the hardness obtained by this method is that of the micro-matrix in the material and does not necessarily correspond to the yield strength of the material on a macroscopic scale. In addition, the means of evaluating ductility, which is important for structural materials, is still under development. Although progress has been made in micromechanical testing techniques, and it will be necessary to irradiate macroscale (or mesoscale) specimens by neutron or proton beam irradiation and carry out mechanical property tests. Even with this limitation in the method, LE ion irradiation is a more cost- and time-efficient method for screening candidate materials, in contrast to the more costly and time demanding high-energy proton irradiations.

Instrumentation performance and its lifetime is often constrained by a prompt radiation dose, and an integrated radiation dose. Having irradiation without activation is an advantage when implementing sensitive or complex instrumentation during on-line and in-situ data acquisition to investigate radiation damage evolution with beam parameters.

**Electron beams** - Passing through matter electrons can interact in many ways with it [11] [12]. A large part of their energy losses will contribute to the total ionizing dose (TID) [11] through the phenomena called soft (interaction with the entire atom) and hard (interaction with one atomic electron) collisions. However, electrons can also collide with nuclei moving them from their position [12]. The result of this, less probable, event is the generation of a vacancy and of an interstitial atom. It worth to note that this is a non-ionizing energy loss (NIEL) [12] [13] and that the process contributes to the displacement damage dose (DDD). Differing from protons or other heavy particles [*12*], electrons (with energy of few MeV) tend to generate isolated defects more than cascade/cluster and their larger penetration in matter allows the irradiation of larger volumes. Electron irradiations performed at different energies (of the order of the MeV) represent a great tool of investigation to reconstruct NIEL curves or to identify the threshold energies of defect formation. We also note that (Figure 1) for electron with energy of about 0.6-3 MeV the NIEL curve can have large variations. While the collision stopping power is almost constant or features low variation, this implies that the TID and the related induced modifications are almost constant too. Therefore, differences in the radiation effects, as a function of the beam energy, can be attributed to the displacement damage. A relevant aspect in performing irradiation is the possibility to record on-line or in-situ data to investigate the radiation effects before any possible recovery processes take place, in particular, temperature changes that would anneal the radiation effects. Such aspect is relevant not only for basic research scopes, but also for applicative reasons since different devices must operate in harsh environments where radiation is present (space, nuclear power plan, large accelerator facility and so on).

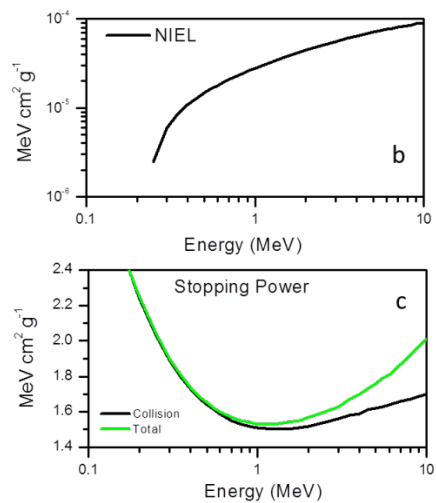

*Figure 1-Top- NIEL curve evaluated with Electrons NIEL Calculator [13] for the Silicon with Exponential form factor and Ed=21 eV; Bottom- Collision and total stopping power for Si data reported in [14].*

**Alternative to mimic gas production** - A characteristic feature of radiation damage by high-intensity proton beams is the formation of helium and hydrogen by nuclear transmutation in the material. When alternative irradiations such as neutron irradiation or low energy heavy ion irradiation, don't provide a similar amount of gas produced by transmutation in the materials as high energy proton, the alternative is to use ion beam facilities with combined beamlines (dual or triple beams) to implant He and H ions. Then, it is possible to investigate the synergistic effects of damage and reproduce transmutation gas production in the material. One drawback is the shallow and non-homogeneous implantation (between 1 to a few micrometers depending on ion energy and material) from low energy ions. Several implantation beam energies are necessary to obtain a homogenous distribution of gas in few microns, increasing significantly the irradiation/implantation time.

## 2.2 Alternative to simulate thermal shock

Testing materials' response to localized thermal shock at this level in a prototypical manner is only possible using an intense particle beam. Alternatives are currently being explored to use intense proton sources, electron sources or laser sources to test thermal shock response more effectively and enable high cycle fatigue studies (> $1 \times 10^7$ loading cycles). The advantage of using electron or laser beams is that we perform testing using prototypical beam loading parameters without activating the specimens.

Currently the HiRadMat beamline at CERN is the only user facility that allows for this intensity but has limitations in the number of consecutive pulses per experiment (~100 pulses maximum), dependence upon the LHC beam operation schedule, and residual activity of the specimens.

## 2.3 Alternative to validate design concept

Testing novel concept design in a prototypical manner is also only possible using an intense beam. Some alternatives are explored with proton and electron high intensity beam.

Electron beams of few MeV have the advantage to simulate the heat deposition in material by adjusting the beam energy in material with reduced risk of activating the tested material. It can be easily used to validate engineering modeling and validate novel concept design.

# 3 Facilities descriptions

As described earlier, it is critical to irradiate Targetry materials under controlled beam parameter to better understand the radiation damage defect creation and the thermal shock that limit beam intercepting devices.

Here is a non-exhaustive list of existing facilities that provide beam available for material studies.

## 3.1 Brookhaven National Laboratory -Brookhaven Linear Isotopes Producer (BLIP)

BLIP is part of Brookhaven Lab's Collider-Accelerator Complex, operating symbiotically with the nuclear physics research program at Brookhaven's Relativistic Heavy Ion Collider (RHIC). The two facilities—one exploring fundamental forces and the building blocks of matter, and the other producing life-saving medical isotopes—make use of alternating pulses of energetic protons and share operating costs and staff. The energetic protons produced by the accelerator can be precisely degraded to a wide range of energies— from 202 MeV down to whatever energy is needed to produce the desired isotopes. The irradiation station for isotope production can be used at some certain level to irradiate material for radiation damage studies.

## 3.2 University of Birmingham

The University of Birmingham hosts an MC40 Cyclotron capable of accelerating protons to energies in the range of 3-39 MeV and currents of a few tens of microamperes. The Cyclotron has been used to study the radiation damage in nuclear materials including RPV steels and tungsten alloys under vacuum [15].

The University is currently constructing an accelerator-driven neutron beam facility at which an intense proton beam (at 2.6 MeV) of up to 30-50mA will be incident on a thick lithium target to generate a neutron beam of up to $3\times10^{12}$ n/cm2/s. The facility is scheduled to be up and running in summer 2022 [16].

For exploiting the proton and neutron beams, the university is currently developing a range of end stations which will enable the irradiation of materials under various conditions relevant to both nuclear fission and fusion applications:

- An aqueous water loop providing pressures and temperatures up to 30 MPa and 650 °C respectively, with a slow strain-rate testing capability, i.e. enabling in-situ irradiation stress corrosion cracking;
- A high temperature vacuum-irradiation mechanical testing capability with temperatures up to 1000 °C.
- A static molten-salt corrosion cell where the sample can be subjected to a controlled stress.

These three end stations are expected to be operational by end of 2022. In the next two years, the University is also planning to develop, jointly with the UK nuclear industry and also international partners, further end stations including the in-situ irradiation and degradation capability in a pressurized gas environment (e.g. supercritical $CO_2$) and also the in-situ irradiation creep capability, among others. There is also the intention to use the two co-located accelerators to develop a dual beam facility which would deliver beam combinations of neutrons with protons and neutrons with helium-4 ions.

### 3.3 TRIUMF

TRIUMF operates several facilities capable of material irradiations using neutrons, 13 – 500 MeV protons, as well as 300 keV – 30 MeV electrons. The TRIUMF-ISAC facility operates two target stations that deliver exotic radioisotope beams to a range of applied science experiments using the ISOL method [17]. The driver beam is a 480 MeV, up to 100 µA, proton beam, generally with a 7 mm FWHM symmetric cross section. The target design consists of a Ta tube containing the target material that operates around 2000 °C, surrounded by a water-cooled heat shield offering an ideal location for a downstream parasitic sample container. Services available for the parasitic irradiations could include thermocouple measurements and independent heating systems.

The small cross section of the driver beam motivates the use of small punch test samples [18] to characterize the mechanical properties of irradiated specimens. FLUKA simulations have been used to evaluate the radiation damage in SPT samples irradiated parasitically in ISAC. An average target irradiation with $2.25 \times 10^{20}$ protons, induce 0.3 FLUKA-DPA [19] (similar to ARC-DPA [20] with 1 ARC-DPA ≈ 3 NRP-DPA [19]) in SS316 (Eth=19 eV), with about 100 appm He/FLUKA-DPA. Per year, samples can accumulate up to 5 times these values by re-coupling the parasitic samples to multiple targets, multi-year programs are possible. About double the DPA rate can be achieved in upstream assemblies that have been tested successfully and irradiate materials directly in the un-scattered proton beam before interacting with the ISOL target. Other locations of the ISAC target heat shield have been used to study the damage in polymers at tens to hundreds of MGy.

Up to 72 samples can be irradiated simultaneously in the proton beam. The 2021 ISAC pilot parasitic irradiation program has irradiated 144 SPT samples (Ø4 mm x 0.25 mm) from 5 different materials: three high-entropy alloys (HEA), SS316, additive manufactured (AM) AlSiMg alloy and Al6061. The samples have accumulated up to 1.4 FLUKA-DPA = 4.7 NRP-DPA in the case of the HEAs and SS316 and 0.2 FLUKA-

DPA = 0.7 NRP-DPA for AM AlSiMg and Al6061. SPT testing of the samples is scheduled for the first semester of 2022.

### 3.4 Irradiation facilities at CERN

Irradiation facilities and test areas operated at the European Organization for Particle Physics (CERN), can be generally divided in two categories.

1. Those related to test of complex systems such as beam intercepting or protection devices, where the objective is the testing of the thermal shock resistance of absorbing blocks as well as to validate integral components.
2. Irradiation facilities supporting equipment installed in accelerator environments (often components and systems based on commercial electronics), large experiments (often custom-made electronics or detector components) and study material damage

It is not possible to detail in short all of the possibilities, so a small selection – relevant for the high power target community – is reported below.

The first category includes the HiRadMat facility [21] a unique high-energy (440 GeV/c) and high intensity pulsed beam facility dedicated to targetry and accelerator components R&D and testing. The facility operates with LHC-like protons or ion beams, with a maximum pulse intensity of $3.5 \times 10^{13}$ protons/pulse (corresponding to roughly 2.5 MJ/pulse), that can be delivered in controlled conditions and monitored with special instrumentations. The power density coupled with the short pulse length (7.2 μs), allows for the testing high power beams on materials and components. Currently, an upgrade study has been launched in order to study the possibility of extracting beams with higher brightness to the facility in order to further explore the material limits for future beam intercepting devices and profiting from the implementation of the LHC Injectors Upgrade (LIU) Project .

The second category also includes several other facilities such as CHARM [22] and IRRAD [23] . The main purpose of CHARM is radiation testing (SEE studies, etc.) of complex electronics equipment systems as well as components in a radiation environment similar to some representative radiation fields that could be find in high energy accelerators. The facility is fed by 24 GeV/c protons impacting on a copper target. IRRAD, instead, employs primary 24 GeV/c protons for direct radiation tests of solid-state and calorimetry detector components or materials, as well as to execute displacement damage and SEE studies on particle detector components.

In the framework of the n_TOF facility spallation target upgrade, a new parasitic, mixed-field, neutron-dominated test area has been recently commissioned, called NEAR. The new irradiation station allows radiation damage studies to be performed in irradiation conditions that are closer to the ones encountered during the operation of particle accelerators; the irradiation tests carried out in the station will be complementary to the standard tests on materials, usually performed with gamma sources. Samples will be exposed to neutron-dominated doses in the MGy/year range, with minimal impact on the n_TOF facility operation. The station has twenty-four irradiation positions, each hosting up to 100 cm$^3$ of sample material.

### 3.5 EMIR&A Network

EMIR&A is the French Federation of Accelerators for Irradiation and Analysis of Molecules and Materials [24]. It is built around several structures and very innovative facilities, that are shared to provide access to ion beams to the largest scientific community possible at the national and international scale. A set of accelerators and instrumentations is made available to users everywhere in the world, after evaluation by an international scientific committee. EMIR&A, itself built on EMIR, extended EMIR from applications of ion and electron irradiation to also include Ion Beam Analysis (IBA). It is a recent step towards an overall national structure for ion and electron beam science and applications such as those which nearby countries have implemented (e.g. the UK National Ion Beam Centre, Surrey, or the Helmholtz-Zentrum at Dresden-Rossendorf in Germany). EMIR&A is also today one of the existing Research Infrastructures of the French Ministry of Higher Education, Research and Innovation, in the field of Material Sciences and Engineering, included in the 2021-2025 roadmap. It aims at connecting the scientific communities that study condensed matter with ion or electron accelerators, to establish new collaborations through the pooling of these instruments, the associated expertise and to animate the scientific community.

EMIR&A includes a total of 15 accelerators installed on 11 platforms distributed over six sites, mainly in the northern half of France (Caen, Orléans) and in the Paris region (Paris, Palaiseau, Orsay, Saclay). Those facilities are owned by many Higher Education institutions, CNRS (National Scientific Research Centre), Universities, Grandes Ecoles and CEA (French Alternative Energies and Atomic Energy Commission), making this structure a worldwide unique portal for such research.

The platforms are organised as a network in order to offer a single portal for hosting external experiments, to create links between them and between users and to promote their development.

In a very synthetic way, a very large panel of materials is studied: nuclear fuel, metals, oxides, ceramics, glasses, minerals, graphite, semiconductors, superconductors, polymers. About a third of the experiments is dedicated to the study of materials for nuclear energy (aging in reactor, new materials for future fission and fusion systems) with the objective of better understanding the behavior and aging of nuclear materials under irradiation from very fundamental aspects to applications. Other studies are interested in the damage mechanisms of materials, structural modifications, modifications of certain properties, mechanical or magnetic for example, or the behavior of solid/solution interfaces under radiolysis. Materials for photovoltaic cells or space are also of growing interest.

## 3.6  J-PARC – Accelerator-Driven-System

To progress the R&D of materials used for Accelerator Driven System (ADS), J-PARC had planned to build a new ADS Target Test Facility called TEF-T [25] [26] [27]. The protons of 0.4-GeV with a power of 250 kW bombard the Lead-Bismuth Eutectic (LBE) and samples in the temperature range of 300-550 °C without the deployment of Minor Actinides (Mas). Although the technical design report of TEF-T was completed, unfortunately, the construction of TEF-T was canceled due to budget conditions. However, J-PARC still wants to build a similar facility enhancing another application with cost reduction. To promote so, J-PARC has an additional plan to use other applications as irradiation facilities for high-energy proton, fusion, and fission facilities. To provide a similar irradiation circumstance of a high-energy proton accelerator, another beam window is planned to be added to enable the proton beam irradiation on the samples in the water. Around the LBE target, except for the window, the irradiation effect is mainly dominated by the neutron. By choosing the sample positions around the target, the user can choose the neutron spectrum to irradiate to their samples. For instance, the neutron spectrum close to the fission reactor can be obtained at a position at a backward angle related to the proton beam direction. All degrees of freedom let us irradiate to multiple-wideband of the neutron spectrum. Also, a new hot cell is planned to be built nearby the new irradiation facility for the PIE. J-PARC does not have any hot-cell for PIE. Due to the law difference in other hot-cell facilities at Japan Atomic Energy Agency (JAEA), where J-PARC is located, the PIE cannot be performed at J-PARC. Furthermore, dismantling the enormous mercury target vessel used for the spallation neutron source at Material Life Science Experimental Facility (MLF) is required, because the tentative storage yard of spent targets has been occupied considerably. The new hot cell will help the R&D of the PIE and dismantle the target vessel for an extended period of operation at MLF.

### 3.7 Paul Scherrer Institut

Since 1998 the SINQ Target Irradiation Program (STIP) has been the unique materials irradiation program performed in spallation targets in the world. 8000+ miniature specimens of various materials have been irradiated in SINQ targets at doses up to 30 dpa (in Fe) and temperatures up to about 600 °C [28] [29].

### 3.8 University of Wisconsin Ion Beam Laboratory

The University of Wisconsin-Madison Ion Beam Laboratory (WIBL) facility uses a 1.7-MV tandem accelerator and features two sources: a TORVIS source used to produce helium and hydrogen ions, to more closely simulate neutron damage in nuclear reactors (using damage rates of ~$10^{-6}$ dpa/s), and a SNICS source capable of producing a wide range of heavy ions (e.g., C, Si, V, Fe, Ni, Nd) used for producing much higher damage rates to simulate end-of-life damage levels in a nuclear reactor (using damage rates of ~$10^{-2}$ dpa/s). The WIBL facility features three post-accelerator beamlines featuring different detectors and equipment to accommodate different ion irradiation experiments; the middle beamline in particular features a large cylindrical chamber (ID: 50 cm, height: 28 cm) which is used to house a novel high-throughput irradiation stage and equipment [30].

With the growing use of combinatorial synthesis methods and high-throughput, automated characterization techniques, ion irradiation facilities need to upgrade their capabilities to be able to accommodate these new material platforms. To address this challenge, a high-throughput ion irradiation system has been developed at WIBL. The system can accommodate large sample arrays of arbitrary geometries and heat samples individually using an infrared laser (beam spot about 1 cm²) to enable high-temperature (up to 800°C), high-throughput ion irradiation while mitigating the risk of annealing the irradiation damage in nearby samples. To control stage motion, laser power, and beam current measurement, among other tasks, the program Chronos was developed which enables automated high-throughput irradiation experiments, with very limited user input. With this new capability, 25 individual irradiations of about 1 cm² each can be performed in about 35 hours using heavy ions up to 200 dpa, making this new capability capable of accommodating large samples arrays such combinatorial thin films [31] and additively printed samples [32].

### 3.9 Michigan Ion Beam Laboratory (MIBL)

The 3 MV Pelletron Tandem accelerator (WOLVERINE) is capable of implantation at energies above 1 MeV and to a maximum energy that depends on the ion charge state, for a maximum terminal voltage of 3 MV. The 1.7 MV Tandetron accelerator (MAIZE) is capable of implantation at energies above 0.5 MeV and to a maximum energy that depends on the ion charge state, for a maximum terminal voltage of 1.7 MV. A wide variety of ions can be produced by a Torvis type source, a duo-plasmatron source, a sputter ion source, SNICS and multi-cathode SNICS sputter sources, and an ECR source, all available on these accelerators. The 400 kV implanter (BLUE) can produce over 40 different ions at energies from 20 to 400 keV. Its end station has stages for implantation at high temperatures down to liquid nitrogen temperature at pressures of $10^{-6}$ to $10^{-7}$ Torr.

The laboratory has been configured to enable dual and triple beam irradiations. The 3 MV accelerator will deliver Fe$^{++}$ up to 9 MeV (~2 µm damage depth in F-M steels) with ion currents up to ~1 µA, yielding a dose rate up to and possibly exceeding $10^{-3}$ dpa/s. Simultaneously, He injection via MAIZE and hydrogen injection via BLUE can be tailored to any implant profile desired, including matching the damage profile.

A dedicated beam line and custom end station provides for conducting irradiation accelerated corrosion experiments in high temperature, high pressure water to simulate the simultaneous effects of irradiation and corrosion. A custom system also provides the capability to conduct irradiation creep with temperature and sample strain measured in a contactless manner.

MIBL houses a ThermoFisher Tecnai G2 F30 TEM interfaced to the 400 kV (NEC) accelerator equipped with a Danfysik Model 921 ion source that can provide a wide range of ions in the 20 keV – 400 keV, single charge state, and 0.8, 1.2 and 1.6 MeV ($2^+$, $3^+$ and $4^+$, respectively) multiply ionized states. The accelerator can deliver specific ion species, such as Kr, Xe and others as required by the users. On a separate beamline, an Alphatross ion source (NEC) has the capability to deliver H and He ions in the energy range of 5-30 keV. The beamline delivery system consists of sophisticated beam tuning devices capable of sending two colinear, but independently controlled ion beams to the TEM with a wide range of fluxes and fluences as required by the user. A state-of-the-art remote-control system allows for beam delivery and TEM operation from a separate Control Room [33] [34] [35] [36] [37].

### 3.10 Grand Accelerateur National d'Ions Lourds, IRRSUD

IRRSUD beamline is one of the four beamlines dedicated to the interdisciplinary physics research of GANIL-SPIRIL2 facilities (Caen, Normandy, France). The platform for welcoming interdisciplinary research is the CIRIL platform [38], which is part of CIMAP [39], a materials science institute. This platform runs the beamlines (D1-HE, D1-ME, IRRSUD and ARIBE) for non-nuclear physics and their equipment, develops new devices requested for experiments, manages the proposal selection process of the local interdisciplinary committee (iPAC) and schedules the interdisciplinary physics beamtimes accepted through iPAC [40], EMIR&A (a network for multi-facility access, see in an earlier paragraph) [41] or RADIATE (an European network for transnational access) [42] committees.

At IRRSUD, available ions ranging from Carbon to Uranium are accelerated by one of the injector cyclotrons (C01 or C02) at energies from 0.3 to 1 MeV/A. These beam energies being below the coulomb barrier. Furthermore, a continuous measurement of the ion flux during irradiation is available coupled to a beam scanning over 5×5 cm² available surfaces. The incident angle for the irradiation can be chosen from normal incidence down to grazing angle of 0.2 degrees. Sample temperature is also a variable parameter from 8 K using a cryogenic holder up to 600 °C (all materials) or even 1200 °C (only nuclear materials) because of two high temperature furnaces. For in-situ and online characterizations, analytical setups have been developed such as :

- absorption spectroscopy in the vibrational range 8000-400 $cm^{-1}$ by FTIR or the electronic range 200-900nm; [43] [44] [45]
- emission spectroscopy by ion luminescence; [46]
- measurement of gas release by FTIR or mass spectrometry; [47]
- structural or microstructural analysis by XRD and grazing configuration; [48]
- sputtering analysis through time of flight or catcher techniques [49].

User's instrumentations can also be mounted on the beamline as already performed for an in-situ UHV AFM [50] or for in-situ tensile test on fibers [51].

### 3.11 Low energy heavy ion beam capacity in Japan

Several facilities are available in Japan: the High Fluence Irradiation Facility of the University of Tokyo (HIT) [52], which is located next to J-PARC, the DuET (Dual-Beam Facility for Energy Science and Technology) [53] at Kyoto University and the TIARA [54] at National Institutes for Quantum Science and

Technology (QST) which are combined ion beam irradiation facilities. These facilities have multiple beamlines that can irradiate material specimens simultaneously.

## 3.12 SIRIUS (Système d'IRradiation pour l'Innovation et les Utilisations Scientifiques)

It is a Pelletron electron accelerator produced by NEC [55], it can produce electron beams with energies from 150 keV to 2.5 MeV and current from 1 nA-50 µA. The accelerator has two beam lines, operating in vacuum ($5 \times 10^{-8}$ mbars) to prevent the degradation of the beam (energy and current) before its arrival on the sample surface. Being equipped with different irradiation cells, Sirius can perform irradiations at temperatures ranging from 600 to 20 K (5 K under development) and in different atmospheres. Furthermore, some irradiation cells are equipped with steps motors that move the sample holder with respect to the electron beam allowing the irradiation of surface up to $180 \times 170$ mm (Grande Surface irradiation cell). The large range of currents allows irradiate with very different flux (dependent on the irradiation cell). The higher values are of about $5.5 \times 10^{13}$ e× $cm^{-2} \times s^{-1}$, allowing to reach fluences of about $4.7 \times 10^{18}$ e×$cm^{-2}$ and dose of about 1 GGy in one day.

## 3.13 Irradiation station capability at Fermilab

The Fermi National Accelerator Laboratory (Fermilab) does not currently have ready capability to do in-beam thermal shock testing. Fermilab has explored using the A2D2 electron linac at I-ARC as well as AP0 line for target testing and more specifically for thermal shock testing and plans to continue pursuing this effort allowed by limited funding. If successful, an electron beam or a proton beam thermal shock testing facility would be a unique and beneficial HPT R&D capability at FNAL. A2D2 could potentially perform accelerated fatigue testing using prototypical beam loading parameters without activating the specimens.

## 3.14 Facility overview

| Facility | Location | Energy range | Intensity range |
| --- | --- | --- | --- |
| BLIP – Brookhaven Linear Isotope Producer | Brookhaven National Laboratory - USA | Up to 200 MeV | 50-165 µA |
| MC40 | University of Birmingham - UK | 2.7-38 MeV | pA to 10's of µA |
| Dynamitron | University of Birmingham - UK | 3 MeV | 1 mA |
| Hyperion | University of Birmingham - UK | 2.6 MeV | 30 mA |
| TRIUMF | Canada | 13-500 MeV | Up to 100 µA |
| MIBL – Maize* for ion implantation | University of Michigan - USA | Up to 2.9 MeV | Up to 2 µA |
| TEF-T | J-PARC - Japan | 400 MeV | 0.6 mA |

*Table 1 – Proton irradiation facilities * for dual and triple beam configuration*

| Facility | Location | Energy range | Fluence or intensity |
| --- | --- | --- | --- |
| PSI* | Switzerland | Up to 550 MeV | Up to $5 \times 10^{25}$ p/m2<br>Up to $1 \times 10^{26}$ n/m² |
| Hyperion | University of Birmingham - UK | | Goal – $3 \times 10^{12}$ n/cm²/s |
| TRIUMF | Canada | | $2 \times 10^{16}$ n/cm²/s |

*Table 2 – Neutron irradiation facilities  * Spallation target*

| Facility | Location | Energy range | Intensity range | Ion Mass range |
|---|---|---|---|---|
| BLIP – Brookhaven Linear Isotope Producer | Brookhaven National Laboratory - USA | 2-28 MeV/A | | 1-197 |
| MC40 | University of Birmingham - UK | 8-50 MeV | | 1-4 |
| WIBL | University of Wisconsin – Madison - USA | 2 MeV | Up to 100 $\mu$A | 1-60+ |
| MIBL – Wolverine* for damage defects | University of Michigan - USA | 1-9 MeV | 500 nA | 1-26 |
| MIBL – Maize* for ion implantation | University of Michigan - USA | Up to 4.5 MeV | Up to 1 $\mu$A | 1-4 |
| MIBL – Blue for ion implantation | University of Michigan - USA | 20-400 keV | Up to 10's of $\mu$A | 1-83 |
| HIT | University of Tokyo - Japan | 0.4-4 MeV | Up to ~ 1 $\mu$A | 1-183 |
| DuET (dual beam) | Kyoto University - Japan | 1-5.1 MeV | 1 $\mu$A-1 mA | 1-197 |
| IRRSUD | GANIL-France | Up to 1 MeV/A | Up to 3 $\mu$Ae | 12-208 |

*Table 3 – Low energy heavy ion irradiation facilities * for triple beam configuration*

| Facility | Location | Energy range | Intensity range |
|---|---|---|---|
| TRIUMF | Canada | 300 keV – 30 MeV | Up to 10 mA |
| SIRIUS | Ecole Polytechnique - France | 150 keV – 2.5 MeV | 10 nA – 50 $\mu$A |

*Table 4 – Electron irradiation facilities*

# 4 Post-Irradiation Examination capability at Facilities

## 4.1 BNL-BLIP

The array of post-radiation capabilities at BNL link campaigns at its irradiation facilities (200 MeV Linac/BLIP and Tandem van de Graaff) with characterization of the effects in both the microstructure and the macrostructure. The Brookhaven Linac Isotope Producer (BLIP) operations have been ongoing for over fifty years for isotope production augmented with proton and spallation-based fast neutron irradiation studies of particle accelerator and nuclear materials. The program has maintained the handling and characterization capabilities of highly radioactive materials. Specifically, capabilities within the BNL hot cell laboratory, essential for nuclear material studies, include photon spectra and isotopic analysis using high-sensitivity detectors, radioactivity measurements and high precision weight loss or gain assessment.

For microscopic characterization, post-irradiation characterization of heavily irradiated materials have relied over the years on the high energy X-rays of its synchrotrons, NSLS (now decommissioned) and now NSLS-II. High energy X-rays at the BNL synchrotrons, and the XPD (and future HEX that is currently under construction) beamlines at NSLS II, offer a path in establishing this important connection between micro-scale effects and physical properties of novel materials exposed to high radiation fluxes. Multiple X-ray techniques – XRD, EDXRD, SAXRD, XAS, X-ray Tomography (XPD, HEX), X-ray Imaging (HEX beamline currently under construction) have been commissioned and utilized extensively to-date to micro-characterize proton and fast neutron irradiated materials for several accelerator and reactor initiatives. Such capabilities include in-situ complex state of stress and various environments.

Characterization of the microstructure under extreme temperatures (currently on unirradiated materials) is provided by the Center of Functional Nanomaterials (CFN) for electron microscopy (TEM, SEM, EDS), thermal analysis (DSC, TGA). In coordination with CFN, plans are being evaluated to enhance the hot cell laboratory (Target Processing Laboratory) with electron microscopy which will, in combination with the micro-characterization techniques at NSLS-II based on high-energy X-rays provide a complete PIE. NSLS-II has developed procedures, including robotics, that allow it to work with radioactive materials with dose rates up to 100 mrem/h at the XPD beamline. Furthermore, BNL is working with DOE-Nuclear Energy on a concept for a beamline with an end station separated from the ring building with controlled access and special capabilities to receive and study radioactive materials with higher dose limits than at any synchrotron beam line within the DOE complex. The initial scope of this special beam line facility will be on structural analysis and tomography but will have the ability to be upgraded to include chemical imaging. In the very near future an intensity upgrade will be under way at BNL to increase the 200 MeV LINAC up to 320 µA peak current it can deliver after its upgrade (peak current achieved to-date 200+ µA) for usable fast neutron spectra for fission and fusion materials.

## 4.2 Michigan Center for Microstructure Characterization (MC)²

(MC)² houses state-of-the-art equipment, including aberration corrected transmission electron microscopes, dual beam focused ion beam / scanning electron microscopes, an x-ray photo-electron spectrometer, a tribo-indenter, an atomic force microscope, and an atom probe tomography instrument. In particular, the laboratory contains the following instruments: Philips XL30 FEG SEM, FEI Quanta 3D e-SEM/FIB, FEI Nova 200 Nanolab SEM/FIB, FEI Helios 650 Nanolab SEM/FIB, JEOL 2010F Analytical Electron Microscope, JEOL 2100F Probe-corrected Electron Microscope, JEOL 3011 High Resolution Electron Microscope, JEOL 3100R05 Double Cs Corrected TEM/STEM, Scanning electron microscopy (SEM), focused ion beam (FIB) milling and imaging, X-ray energy dispersive spectrometry (XEDS), Electron backscattered diffraction (EBSD), cryo electron microscopy, transmission electron microscopy (TEM), including diffraction imaging, high resolution (HREM), scanning (STEM), and aberration-corrected microscopy, in-situ electron microscopy (straining, heating, indentation), electron energy loss spectrometry (EELS), atom probe microscopy (APM), atomic force microscopy (AFM), x-ray photoelectron spectroscopy (XPS), tribo/pico-indentation.

## 4.3 PIE capability at TRIUMF

Additionally, routine replacement of ISOL targets in dedicated hot cells allows for a flexible irradiation program and PIE capabilities. PIE currently includes Small Punch Tests (SPT) and SEM, a TEM is being installed and other characterization techniques are under consideration.

## 4.4 PIE capability at Oak Ridge National Laboratory

Oak Ridge National Laboratory (ONRL) has several world-leading facilities with advanced characterization capabilities for post-irradiation examination (PIE) of radioactive materials. Two general options exist at ORNL depending on the dose rates of the specimens.

Specimens with relatively low dose rates (<0.1 mSv/h at 30 cm) may be characterized at the Low Activation Materials Development Laboratory (LAMDA) facilities at ORNL, though exceptions can be made for slightly higher dose rate specimens on a case-by-case basis. The LAMDA user center is a collection of laboratory facilities at ORNL with advanced mechanical properties and microstructure characterization capabilities. Several different size load frames equipped with digital image correlation (DIC) are available for testing materials at a wide range of temperatures and environments. In-situ scanning electron microscope (SEM) tensile testing accompanied with electron backscattered diffraction (EBSD) analysis is available for deformation mechanism studies. Micro- and nano-hardness testing machines are available. The LAMDA facilities also host a wide variety of scanning electron microscopes (SEM), transmission electron microscopes (TEM), and atom probes that are capable of most advanced characterization techniques currently available, including bright field TEM (BF-TEM), scanning TEM (STEM), high-resolution TEM (HR-TEM), electron energy-loss spectroscopy (EELS), and atom probe tomography. Thermal desorption spectrometry is also available for H and He gas measurements of irradiated specimens.

Several hot cell facilities are available at ORNL for testing and characterization of specimens with high dose rates (>0.2 Sv/h at 30 cm). Multiple load frames with different capacities are available for tensile and fracture toughness testing over a wide range of temperatures. Other mechanical testing available include microhardness, automated ball indentation, and Charpy impact toughness testing. In-cell space is also usually available for development of unique capabilities for specific experimental programs.

### 4.5 PIE capability in Japan with associated irradiation stations

At the Multi-Quantum Beam High Voltage Electron Microscope Laboratory [56], Hokkaido University, a 1300 keV high-voltage electron microscope coupled with two ion accelerators enables in-situ observation of defect behavior under helium and electron irradiation.

### 4.6 PIE capability at IRRSUD Facility

Besides the in-situ and online characterization techniques, many post irradiation experiments are also available through collaborations at CIMAP. A list including but not limited to is: high-resolution XRD, dual-beam FIB/SEM, TEM, AFM, Raman/FTIR/Uv-vis spectroscopy, resistivity at RT and many others.

### 4.7 PIE capability at EMIR&A network

The facilities are complementary by the nature and energies of the accelerated particles, and the associated IBA and in situ material characterization instrumentation such as Transmission Electron Microscopy (TEM), Raman spectroscopy, X-Ray Diffraction (XRD), infrared spectroscopy, or optical absorption [21].

### 4.8 PIE Capability at PSI

A full spectrum of PIE facilities is available for mechanical testing and microstructural analysis of irradiated specimens. Specimens of activity up to ~10 GBq can be mechanically tested at temperatures between 20 and 600 °C. Specimens of activity up to ~100 MBq can be analyzed with TEM, SEM, APT, PAS, SANS, synchrotron X-ray absorption spectroscopy. In addition, samples can be extracted from spent parts from nuclear power plants, spallation targets and accelerator components.

### 4.9 PIE capabilities at Fermilab

Very limited PIE capabilities (Tensile machine, fatigue testing machine, digital microscopy, dilatometer, etc) exist at Fermilab and only non-activated material or very low activated (< 40 mrem/h at 1 ft) can be tested under specific conditions.

To support LBNF project, the Target Systems Integration Building (TSIB) is under development as a General Plant Project (GPP). Fully integrated to this building the High Power Targetry Laboratory will have the capability to support:

- Radiation damage studies of irradiated materials – Materials that have been previously irradiated will be studied and tested to determine the effects of accelerator-based irradiation at different conditions upon physical and mechanical properties. See the TSIB HPT Lab Material Workflow Plan for details on sources of irradiated materials, associated levels of activity, and plans for safe handling of active materials in the HPT Lab.
- Materials science studies of non-active materials – Targetry materials and materials under consideration for use as targetry material will be studied and tested to establish baselines and evaluate their potential.
- Development of novel materials for use as targetry material and associated fabrication and production technologies.

This project is essential to address the needs and concerns for future generation accelerator.

### 4.10 PIE capabilities at PNNL

The Pacific Northwest National Laboratory (PNNL) maintains extensive PIE capabilities, housed primarily in two facilities. The Radiochemical Processing Laboratory (RPL) is a Hazard Category 2 nuclear facility capable of studying irradiated fuel and high-activity structural materials, and the Materials Science and Technology Laboratory (MSTL) is a radiological facility that focuses on lower-activity structural materials. PNNL can accept a variety of shipping containers ranging from commercial spent fuel casks such as the NAC-LWT, smaller Type B casks such as the GE-2000 and 10-160B, as well as Type A containers in all shapes and sizes. Within the RPL, there are several sets of hot cells, ranging from the High Level Radiochemistry Facility (HLRF) with three high-density concrete hot cells with 1.2 m thick walls ranging in size from $1.8 \times 2.6 \times 5.2$ m to $4.5 \times 2.6 \times 5.2$ m. All three cells are interconnected for easy sample transfers. Existing capabilities in HLRF include mechanical rod puncture, plenum gas collection and online gas analysis via gamma energy analysis and mass spectrometry, internal rod pressure measurement, and gamma spectroscopy. After receipt, samples are size-reduced in HLRF and distributed to other laboratories within RPL for subsequent analysis. A bank of six interconnected 1 m concrete and steel hot cells with internal dimensions of $1.7 \times 1.7 \times 5.2$ m comprise the Shielded Analytical Laboratory (SAL) at RPL. The SAL hot cells can be used for precision sample preparation, chemical treatment or separations, or analytical procedures. In addition to HLRF and SAL, there are seven standalone modular hot cells with 30 cm steel shielding in a variety of sizes. Each has large access doors for easy equipment installation and removal, and they are configured as project needs dictate. Testing capabilities in the modular hot cells include mechanical properties using an Instron 8800 load frame. An electrospark discharge machining (EDM) instrument has been purchased and will be installed in one of the modular hot cells in 2023 to provide precision machining of irradiated materials. There are extensive microscopy capabilities in RPL, including a fully-automated and remote-operated optical microscope, a focused ion beam (FIB) SEM, another SEM with EDS/WDS/EBSD capability, a JEOL GrandARM 300F aberration-corrected STEM, and an atomic force microscope configured for topography, nanohardness, thermal conductivity and magnetic property measurements. An ion mill was recently installed for optimized preparation of surfaces for EBSD analysis. A Cameca LEAP 5000 atom probe tomography instrument was recently installed. The RPL includes a full suite of analytical chemistry capabilities including high-resolution hydrogen isotope mass spectrometry, high precision (ppt) He isotope mass spectrometry, TIMS, ICP-OES, ICP-MS, gas mass spectrometry, ion chromatography, NMR spectroscopy and Raman spectrophotometry. There are also surface science capabilities including Auger electron spectroscopy, x-ray photoelectron spectroscopy, secondary ion mass spectrometry and Fourier transform infrared spectroscopy. In addition, expertise exists in the RPL with respect to flux wire manufacturing and post-irradiation gamma spectroscopy for determining flux, fluence, and dpa. At the Materials Science and Technology Laboratory, there are complementary microscopy capabilities including optical, a FIB/SEM, SEM with EDS/WDS/EBSD capability, and a JEOL ARM 200F aberration-corrected STEM. There are also load frames for mechanical property testing of activated structural materials including tensile and fracture toughness measurements, Vickers microhardness testing capability, and autoclaves for corrosion and stress corrosion cracking experiments.

# 5 Conclusion

As presented in this document, few facilities exist to support several aspects of the R&D program to ensure reliable operation of future target facilities:

- Irradiation under controlled parameters for better understanding of radiation damage behavior in selected material for beam intercepting devices
- Develop alternative methods to high energy proton beam to reduce cost and time of R&D cycle
- Collect experimental data on irradiated material to support advanced modeling on material development
- Test prototypes under similar operational conditions in a safer environment (without activating material) and validate engineering modeling for novel concept design

Nevertheless, each facility has limited access to users and Post-Irradiation Examination may not meet the need of specific test related to accelerator Targetry development. Some PIE capabilities are limited or are not existing near some irradiation stations, leading to transport activated materials to other capable facilities, increasing cost and time of the R&D cycle.

Other facilities are only dedicated for operation but could be developed at lower cost to offer testing capabilities or propose an irradiation station as parasite mode to limit interference with operation. Fermilab is exploring existing facilities as well as future facilities such as PIP II to assess irradiation station implantation.

In the next 10 years a priority should be focus on the development and the validation of the alternative beams to simulate high energy proton irradiation. If successfully proven, those methods could potentially reduce cost and time to narrow material selection, to extrapolate radiation damage behavior with advanced modeling.

The target community has been collaborating through the High Power Targetry Workshop since 2003 and through the RaDIATE collaboration, led by Fermilab, since 2013. Except the EMIR&A network in France that regroups a set of 15 accelerators and instrumentations made available for users to focus on condensed matter studies with ion or electron beams. This network, based on a single portal, facilitates experiments and create a strong synergy between the users and the facilities.

A similar network within the NSUF mission, exists in US to support nuclear science and technology by providing nuclear energy researchers with access to world-class capabilities at no cost to the researchers. But HEP researchers don't have access to this network and no similar platform with irradiation facilities and associated PIE exists in US to support HEP R&D program. Such national or international collaborations and network will be essential in the next 10 years to expand our knowledge on high power targets.

Accelerator facilities and GARD (DOE HEP General Accelerator R&D) will need to invest more in the target R&D to develop irradiation capabilities and get a better understanding beam effect in material (i.e. radiation damage, thermal shock and fatigue), validate the target modeling and advanced novel concept to the needed level for high power accelerators.